\documentclass{emulateapj} 
 
\newcommand{\lsim}{\raisebox{-0.13cm}{~\shortstack{$<$ \\[-0.07cm] $\sim$}}~}
\newcommand{\gsim}{\raisebox{-0.13cm}{~\shortstack{$>$ \\[-0.07cm] $\sim$}}~}

\shorttitle{Linking stellar mass and star formation} 
\shortauthors{K.I. Caputi et al.} 
 
\begin{document} 
 
 
\title{Linking stellar mass and star formation in Spitzer/MIPS 24$\,\mu \lowercase{\rm m}$ galaxies} 
 
 
\author{K.I. \ Caputi\altaffilmark{1}, 
H. \ Dole\altaffilmark{1},
G. \ Lagache\altaffilmark{1},
R.J. \ McLure\altaffilmark{2}, 
J-L. \ Puget\altaffilmark{1}, 
G.H.\ Rieke\altaffilmark{3}, 
J.S. \ Dunlop\altaffilmark{2},
E.\ Le Floc'h\altaffilmark{3},
C. \ Papovich\altaffilmark{3},
P.G. P\'erez-Gonz\'alez\altaffilmark{3}
} 
\altaffiltext{1} {Institut d'Astrophysique Spatiale, b\^at. 121, 
Universit\'e Paris-Sud, F-91405 Orsay Cedex, FRANCE}
\altaffiltext{2} {Institute for Astronomy, University of Edinburgh, Royal Observatory, Blackford Hill, 
EH9 3HJ, Edinburgh, UK}  
\altaffiltext{3} {Steward Observatory, University of Arizona, 933 N 
Cherry Ave, Tucson, AZ 85721, USA}


 
\begin{abstract} 
  We present  deep $K_s < 21.5$ (Vega) identifications, redshifts and stellar masses for  most of the sources composing the bulk of the  $\rm 24 \,\mu m$ background in the GOODS/CDFS.  Our identified sample consists of 747 {\em Spitzer}/MIPS $\rm 24 \,\mu m$ objects, and includes $\sim 94$\% of all the $24 \, \mu{\rm m}$ sources in the GOODS-South field which have fluxes $S_\nu(\rm 24 \,\mu m)> 83 \, \mu {\rm Jy}$ (the $\sim 80$\% completeness limit of the {\em Spitzer}/GTO $\rm 24 \,\mu m$ catalog). 36\% of our galaxies have spectroscopic redshifts (mostly at $z<1.5$) and the remaining ones have photometric redshifts of very good quality, with a median of   $|dz|=|z_{spec}-z_{phot}|/(1+z_{spec}) = 0.02$. We find that MIPS $\rm 24 \mu m$ galaxies span the redshift range $z \sim 0-4$, and that a substantial fraction (28\%)  lies at high redshifts $z \gsim 1.5$. We determine the existence of a bump in the redshift distribution at $z \sim 1.9$, indicating the presence of a significant population of galaxies with PAH  emission at these redshifts.  The $24 \,\mu {\rm m}$ galaxy population ranges from sources with intermediate luminosities ($10^{10}\, L_{\odot}<L_{IR}<10^{11}\, L_{\odot}$) and low-to-intermediate assembled stellar masses ($10^{9} \, M_{\odot} \lsim M \lsim 10^{11}  \, M_{\odot}$) at $z \lsim 0.8$, to massive ($M \gsim 10^{11}  \, M_{\odot}$) hyper-luminous galaxies ($L_{IR}>10^{12}\, L_{\odot}$) at redshifts $z \sim 2-3$.  Massive star-forming galaxies at redshifts $2 \lsim z \lsim 3$ are  characterized by very high star-formation rates ($SFR>500 M_\odot/{\rm yr}$), and some of them are able to construct a mass of $\approx 10^{10}  \, M_{\odot} - 10^{11}  \, M_{\odot}$ in  a single  burst lifetime ($\sim$ 0.01 Gyr - 0.1 Gyr). At lower redshifts $z \lsim 2$, massive star-forming galaxies are also present, but  appear to be building their stars on long timescales, either quiescently or in multiple modest burst-like episodes. At redshifts  $z \sim 1-2$, the ability of the burst-like mode to produce entire galaxies in a single event is limited to some lower ($M \lsim 7 \times 10^{10} M_\odot$) mass systems, and it is basically negligible at $z \lsim 1$. Our results support a scenario where star-formation activity is differential with assembled stellar mass and redshift, and where the relative importance of the burst-like mode proceeds in a down-sizing  way from high to low redshifts.   
\end{abstract} 
 
 
 
 
%
\section{Introduction}   
 
 Infrared (IR) surveys provide an unbiased way of studying the star-formation history of the Universe.  The {\em Spitzer Space Telescope} (Werner et al. 2004), with its unprecedented sensitivity, is revolutionizing our understanding of IR galaxy evolution by obtaining very deep IR maps.  The previous space IR facilities,  the  Infrared Astronomical Satellite (IRAS) and the Infrared Space Observatory (ISO), set the first constraints on mid and far-IR galaxy evolution at $z \lsim 1$ (see Genzel \& Cesarsky 2000 and  Franceschini et al. 2001, for a review). Deep {\em Spitzer} images offer the possibility of exploring star-formation in the high-redshift Universe, covering the gap between ISO sources and the still limited quantity of known submillimetre sources at redshifts $z  \sim 2-3$ (e.g. Scott et al. 2002; Borys et al. 2003; Chapman et al. 2003).

 The study of IR sources at high redshifts is useful to put constraints on the early stages of star and galaxy formation. An important cosmological issue is the determination of what kind of galaxies host star-formation activity at different redshifts. At  $z \sim 1$, a mixture of star-forming objects is found, from massive spiral galaxies to on-going starbursts (Franceschini et al. 2003; Bell et al. 2005). In addition, it is known that star-formation can proceed on different timescales, from bursts produced in short time periods (typically 0.01 Gyr to 0.1 Gyr) to sources with extended periods of quiescent activity. The study of the evolution of the full range of star-forming galaxies should allow reconstruction of the star-formation history of the Universe.

 Several empirical approaches based on IRAS/ISO surveys have predicted the existence of a substantial population of  mid-IR galaxies at high redshifts. Very recently,  based on the modelling of the number counts of {\em Spitzer} $\rm 24\,\mu m$ galaxies, Lagache et al. (2004) concluded that the Policyclic Aromatic Hydrocarbon (PAH) emission characteristic  of  the interstellar medium (D\'esert, Boulanger \& Puget 1990) should remain observable in galaxies up to $z \sim 2.5$. The main PAH spectral features are located at rest-frame wavelengths $\rm \lambda_{rf}=3.3, 6.2, 7.7, 8.6, 11.3, 12.7, 16.3$ and $\rm 17 \, \mu m$. If present, these emission lines should enter the $\rm 24\,\mu m$ filter at redshifts $z \sim 6.3, 2.9, 2.1, 1.8, 1.1, 0.9, 0.5$ and $0.4$, respectively.  Thus, if PAH molecules  already exist in high-redshift galaxies, they should appear in the $\rm 24\,\mu m$ passband producing the selection of a substantial population of sources at redshifts  $z \gsim 1$ (Papovich et al.~2004).

  In this work we present $K_s$-band identifications, redshifts and stellar mass estimates for most of  the sources composing the $\rm 24\, \mu m $  background down to faint fluxes in $\sim$ 131 arcmin$^2$ of the Great Observatories Origins Deep Survey (GOODS)/ Chandra Deep Field South (CDFS), achieving 94\%  identification completeness for sources with  flux $S_\nu[24 \, \mu m] > 83 \, \mu {\rm Jy}$.  The mostly complete  redshift identification of {\em Spitzer}  $\rm 24\, \mu m $ sources with magnitudes $R<24$ (Vega) at $z \lsim 1$  has already been achieved by Le Floc'h et al.~(2005) over an extended area of  the CDFS. Evidence  of  luminous infrared galaxies up to redshift $z \sim 2.5$ has been reported by different authors (Le Floc'h et al.~2004; Lonsdale et al.~2004; Papovich et al. 2005).   The evolution of the mid-IR luminosity function and the derived SFR densities have been explored by Le Floc'h et al.~(2005) and P\'erez Gonz\'alez et al. (2005).   The present work, albeit based on the study of a smaller area, complements the former as it  exploits the unique quality of the GOODS datasets to obtain an almost complete identification  of  the sources composing  the mid-IR background and to characterize some of their most important properties from low to high redshifts ($z \approx 0$ to $z \approx 3-4$).
  
   The organization of this paper is as follows. In Section \ref{sec_data}, we summarize the datasets and catalogs used to construct the source sample analyzed in the present study. In Section \ref{sec_redsh}, we present the redshift distribution of the $\rm 24\, \mu m $ galaxy population in the GOODS/CDFS. In Section \ref{sec_stmass}, we give estimates of their characteristic  stellar masses. In Section \ref{sec_stform}, we study the evolution of IR luminosities and stellar masses with redshift and put constraints on the star-formation rates (SFR) and timescales for star-formation activity in different-mass galaxies.  
 Finally, in Section \ref{sec_disc}, we summarize our results and present some concluding remarks. We adopt throughout a cosmology with $\rm H_o=70 \,{\rm km \, s^{-1} Mpc^{-1}}$, $\rm \Omega_M=0.3$ and $\rm
\Omega_\Lambda=0.7$.
%
\section{$K_s$-band counterparts to MIPS $\rm 24 \, \mu \lowercase{m} $ sources} 
\label{sec_data} 

\subsection{The data samples and redshift estimates}

 Deep $\rm 24\, \mu m $ observations of $\rm \sim 2000 \, arcmin^2$ of the CDFS have been carried out with the  Multiband Imaging Photometer for {\em Spitzer} (MIPS; Rieke et al. 2004),  as part of the Guaranteed Time Observers (GTO) program. The data reduction, extraction of sources and photometry measurements are explained in detail in 
 Papovich et al. (2004). The source catalog achieves 80\% completeness at a flux limit of $ S_\nu\sim 83 \,\mu$Jy and the fraction of spurious
 sources at that limit is determined to be $<$10\%.  The $\rm 24\, \mu m $ source density in the CDFS is N($ S_\nu > 83 \,\mu$Jy)$\rm =4.5 \, arcmin^{-2}$.  Above a flux of $S_\nu\sim 60 \,\mu$Jy, resolved {\em Spitzer} sources account for $\sim$ 70\% of the   $\rm 24\, \mu m $  cosmic infrared background (Papovich et al. 2004; Lagache et al. 2004).

 The GOODS project \cite[]{giav04} has provided deep  multiwavelength data for $\rm \sim 160 \,  arcmin^2$ of the CDFS.
   The data products are released to the astronomical community in a fully
 reduced mode.  As part of the European Southern Observatory GOODS Imaging Survey program (ESO GOODS/EIS; Vandame et al., in preparation), deep J and
 $K_s$-band images taken with the Infrared Spectrometer and Array Camera (ISAAC) on the Very Large Telescope (VLT) have been made public for $\rm \sim 131 \, arcmin^2$ of the GOODS/CDFS. This is the area analysed in the present work. The ESO/EIS imaging in this field is
 complemented by deep imaging data in the $B$, $V$, $I_{775}$ and $z_{850}$ bands taken with the Advanced Camera for Surveys (ACS) on board the Hubble Space Telescope (HST) and by $\rm 3.6\, \mu m, 4.5\, \mu m, 5.8\, \mu m$ and $\rm 8.0\, \mu m$ data taken with the Infrared Array Camera (IRAC) \cite[]{fazio04} on board {\em Spitzer}. 
 
  A catalog of $K_s<21.5$ (Vega) sources has been selected in the GOODS/EIS CDFS. Photometric redshifts based on up to eight broad-bands ($B V I_{775} z_{850}  J K_s$ and IRAC $\rm 3.6 \, \mu m$ and $\rm 4.5\, \mu m$ bands) have been obtained for the whole sample, using the publicly available multiwavelength data and the public code HYPERZ (Bolzonella, Miralles \& Pell\'o, 2000) with the GISSEL98 spectral energy distribution (SED) template library (Bruzual \& Charlot, 1993). Dust-corrections have been taken into account through the convolution of the SED templates with the Calzetti et al. (2000) reddening law.
  
\begin{figure} 
\plotone{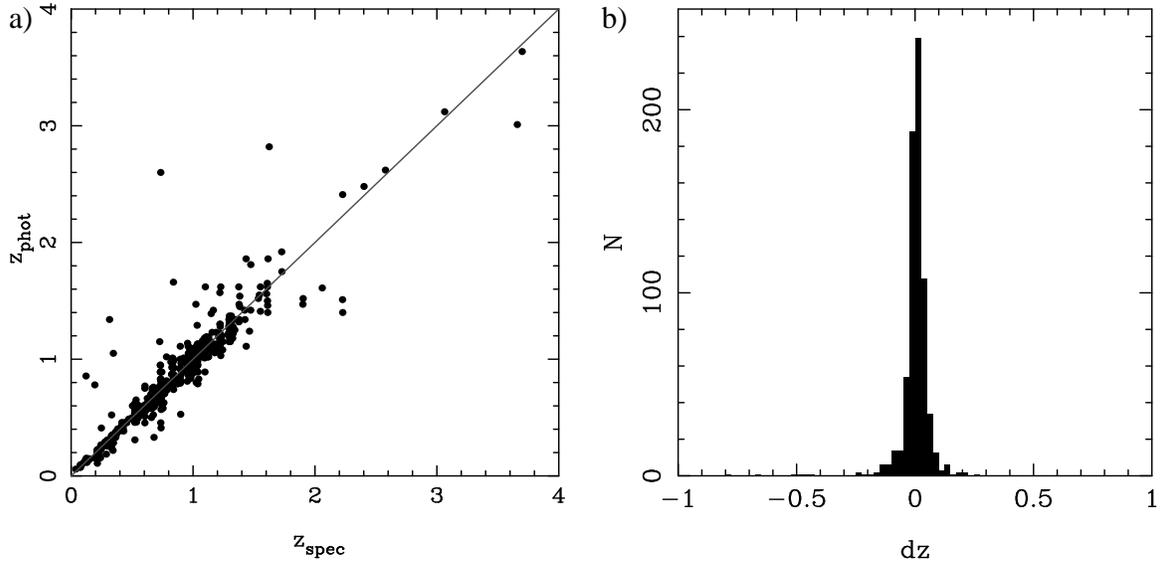} 
\caption{a) Photometric versus spectroscopic redshifts for $K_s<21.5$ galaxies in the GOODS/CDFS. b) Histogram of the relative errors $dz=(z_{spec}-z_{phot})/(1+z_{spec})$. The rms of this distribution is 0.03.
 \label{zqual}} 
\end{figure}

     Although U-band data exist for the CDFS, they have not been included  in the input catalogs for the photometric redshift algorithms as the U-band images have shallower depth and poorer resolution than the ACS and ISAAC images in the same field. However, U-band images have been used to control the HYPERZ output: for every $K_s$-band source with a counterpart in the shallow U-band catalogs, the  resulting photometric redshift was determined as the best-fit value constrained to a maximum redshift $z_{phot}=2$, as higher redshift sources are unlikely to be bright at such short wavelengths. In addition, the public code BPZ (Ben\'{\i}tez 2000) has been used to obtain a second, independent, set of redshift estimates for all the $K_s<21.5$ sources. In the cases of sources with HYPERZ photometric redshifts $z_{phot}>2$ not confirmed by BPZ, we adopted the lower estimates from the BPZ code. In addition,  public COMBO17 photometric redshifts \cite[]{wolf04} have been used to replace the redshift estimates of those sources with  magnitude $R<23.5$ (Vega) and $z<1$ (the most accurate regime for COMBO17 redshifts). Further details on the redshift estimations for the total $K_s<21.5$ sample are given in a separate paper (Caputi et al. 2005b).  Figure~\ref{zqual}a) compares the obtained photometric redshifts with spectroscopic redshifts for all those $K_s<21.5$ sources in the GOODS/EIS CDFS included in different publicly available spectroscopy samples (e.g. Vanzella et al. 2005; Le F\`evre et al. 2004). We observe a very good agreement between  the photometric estimates and the real redshifts in most cases, with a median for the absolute relative errors of   $|dz|=|z_{spec}-z_{phot}|/(1+z_{spec}) = 0.02$.  Figure~\ref{zqual}b) shows the corresponding histogram of relative errors. The rms of the distribution is 0.03. However, in spite of the good accuracy obtained for the photometric redshifts, the incorporation of available spectroscopic redshifts is of much benefit for any study of a galaxy redshift distribution. Thus, in order to maximize the quality of the final redshift catalog, the existing spectroscopic redshifts in the CDFS have been used to replace the photometric estimates of the $K_s<21.5$ sources whenever possible (i.e. 23\% of the total $K_s<21.5$ sample).

\subsection{Cross-correlation of the catalogs}

    We cross-correlated the  $K_s<21.5$ GOODS/CDFS source catalog with the Spitzer/MIPS $\rm 24\, \mu m $ GTO/CDFS catalog  both to investigate and put constraints on  the mid-IR emission of $K_s$-selected galaxies and to identify and characterize the optical/near-IR properties of the {\em Spitzer}/MIPS $\rm 24\, \mu m $ sources in the GOODS South field. This work presents the results of the latter, while a full analysis of  $\rm 24\, \mu m $-detected and non-detected $K_s$-selected galaxies in the GOODS/CDFS will be presented  elsewhere (Caputi et al. 2005c, in preparation).  In addition, the IRAC and MIPS properties of galaxies selected with colours $J-K_s>2.3$  are studied by Papovich et al.~(2005).

\begin{figure} 
\plotone{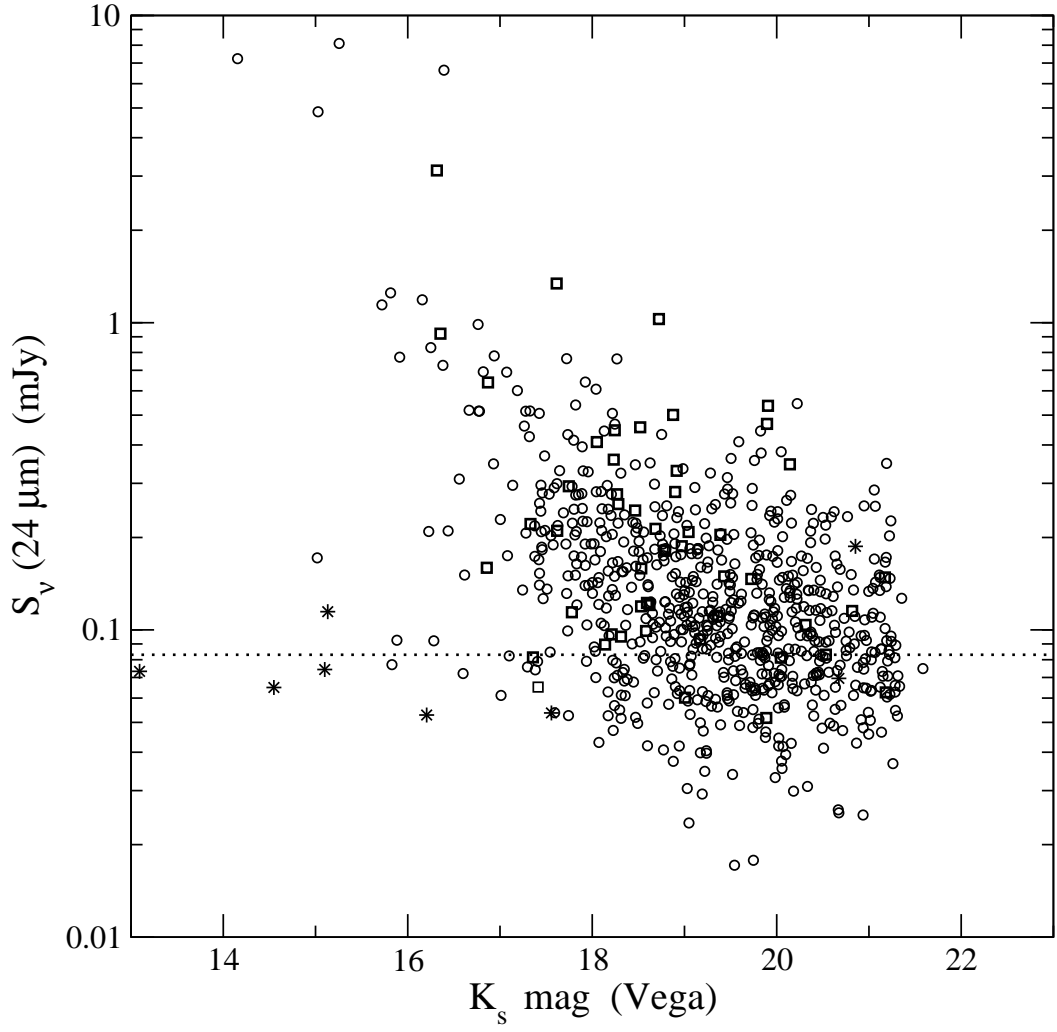} 
\caption{ $\rm 24\, \mu m $ fluxes versus $K_s$ magnitudes for the MIPS $\rm 24\, \mu m $  sources in the GOODS/CDFS associated to a $K_s<21.5$  counterpart. The circles, squares and asterisks  correspond to normal galaxies, active galaxies and galactic stars, respectively. The dotted line indicates the limit of $\sim 80$\% completeness of the 
$\rm 24\, \mu m $ catalog, above which the $K_s$-band identification is $\sim 94$\%.
 \label{24vsKs}} 
\end{figure}

   We looked for $ \rm 24 \,\mu m $  counterparts of the $K_s<21.5$-selected sources in the GOODS/CDFS using a matching radius of $2^{\prime\prime}$, to minimize the number of   multiple identifications. The astrometric accuracy of the MIPS-24$\rm \mu m$  images (e.g. Le Floc'h et al. 2005) allows  the use of such a small matching distance.  We found 812 $K_s<21.5$-selected sources associated with MIPS $\rm 24\, \mu m $  sources within this radius. In 65/812 cases, we found two or more $K_s$-selected sources associated with a same $\rm 24\, \mu m $  source. In these cases we selected the closest counterpart, leaving 747 $\rm  24 \,\mu m $ sources with a $K_s<21.5$ identification. We identified a negligible fraction (11/747) of galactic stars  among these MIPS  $\rm  24 \, \mu m $  sources, and we exclude them from all the following analysis. We determined redshifts for each of the remaining 736  $\rm  24 \, \mu m $ sources. 36\% of them have spectroscopic redshifts and  21\% more have COMBO17 photometric redshifts.  The remaining 43\%  have HYPERZ/BPZ photometric redshifts.  521/747 sources are above the $\rm 24\, \mu m $-catalog  80\% completeness limit, i.e. have $S_\nu > 83 \,\mu{\rm Jy}$.  On the other hand, we find 4  $K_s$-selected objects that could be associated with two or more  $\rm 24 \,\mu m $ sources. All  of these cases correspond to low redshift galaxies, with spectroscopic redshifts $z_{spec}<0.13$. Individual inspection of  both the $K_s$ and  $\rm 24 \, \mu m $ images suggests  the Spitzer/MIPS detection of multiple nodes of mid-IR emission for each of these nearby galaxies.   In Figure~\ref{24vsKs}, we show the total  $\rm 24 \, \mu m $ flux  $S_\nu$ versus $K_s$ magnitude for the 747 MIPS sources with $K_s<21.5$ identifications in the GOODS/CDFS.  In the 4 cases of multiple MIPS sources  associated to a single  $K_s$-band object,   we considered that the  $\rm 24 \, \mu m $ flux  was the sum of the different $\rm  24\, \mu m $ components.  686/747  MIPS  $\rm 24 \, \mu m $  sources in the GOODS/CDFS have  counterparts classified as normal galaxies (circles in Figure~\ref{24vsKs}), while 50/747 are active ones (AGNs or QSOs; squares in Figure~\ref{24vsKs}). The identification  of $K_s$ sources with active galaxies is based on the cross-correlation with the X-ray catalogs available for the CDFS  \cite[]{szo04}. This approach provides only a lower limit to the fraction of active galaxies among the  $\rm 24 \, \mu m $ galaxy population (Alonso-Herrero et al. 2005; Donley et al. 2005).  The cross-correlations of  the $K_s<21.5$ GOODS/CDFS source catalog with the {\em Spitzer}/MIPS $\rm  24 \,\mu m $ GTO/CDFS catalog allows us to achieve a completeness limit of $\sim 94$\% for the identification of MIPS $\rm 24 \, \mu m $  sources with flux $ S_\nu > 83 \,\mu$Jy in the GOODS/CDFS (cf. Table~\ref{tab1}).

  As we mentioned above, our HYPERZ photometric redshift determinations are based on the SED fitting of the galaxies made using the templates in the GISSEL98 library of Bruzual \& Charlot. Also, this SED fitting procedure has been used to determine derived parameters (e.g. estimated stellar mass) for all the galaxies, independently of whether the redshifts were from HYPERZ/BPZ, COMBO17 or spectroscopic.  The  GISSEL98 library is composed of a wide range of synthetic SEDs based on stellar spectra but does not have any optical to near-IR power-law SED. To investigate the impact of this limitation in our redshift distribution of  MIPS $\rm 24\,\mu m$ galaxies, we identified the presence of $\rm 24\,\mu m$ power-law SED galaxies  in the GOODS/CDFS using the catalog of IR-power-law sources constructed by Alonso-Herrero et al.~(2005). We found 20  $\rm 24\,\mu m$ galaxies with  IR-power-law spectra in the GOODS/CDFS. 9 of these power-law galaxies have spectroscopic redshifts, leaving only 11 power-law sources with a HYPERZ/BPZ photometric redshift (and 7/11 are X-ray detected; see Section \ref{sec_stmass}). Consequently, plausible erroneous redshift estimates due to inadequate SED template fitting are reduced to a very minor fraction of the $\rm 24\,\mu m$  sources studied here and, thus, they  should have a  basically negligible impact on all the analysis  in this work.

\subsection{Summary of properties of the $ \rm 24 \,\mu m $ source sample analyzed in this work}

 For clarity, we summarize the properties of the  source sample analyzed in this work:

\begin{itemize}
\item our sample is composed of the 747 $ \rm 24 \,\mu m $ sources in $\rm \sim 131 \, arcmin^2$ of the GOODS/CDFS which have a $K_s<21.5$ (Vega) counterpart. 521/747 sources have a flux above the GTO catalog 80\%  completeness limit, i.e.  $S_\nu(\rm 24 \,\mu m)> 83 \,\mu{\rm Jy}$.  These 521 objects constitute $\sim 94$\%  of all the MIPS $ \rm 24 \,\mu m $ sources with $S_\nu(\rm 24 \,\mu m)> 83 \,\mu{\rm Jy}$ in the GOODS/CDFS.

\item Only 11/747 sources in our sample are identified as galactic stars. 50/747 sources are X-ray classified AGNs or QSOs.  The remaining 686/747 sources are normal galaxies (i.e. no X-ray AGN or QSO). 

\item  We determined redshifts for each of the 736 galaxies in our sample (i.e. all sources except the 11 galactic stars).  36\% of these redshifts are spectroscopic.  21\%  additional redshifts have been  taken from the COMBO17 photometric redshift catalog (only for sources with R-band magnitude $R<23.5$, Vega, and redshifts $z<1$). The photometric redshifts for the remaining 43\% of the sample have been obtained using HYPERZ/BPZ and are based on  broad-band photometry covering from the optical B-band through the  $\rm 4.5\, \mu m$ band.

\end{itemize}
\noindent In this way, our sample allows us to identify and characterize the evolution down to faint fluxes for most of the sources comprising the $ \rm 24 \,\mu m $ background.

   P\'erez-Gonz\'alez et al. (2005) have carried out a study of MIPS $ \rm 24 \,\mu m $ detections using a different photometric redshift technique. Their approach uses empirical SEDs rather than synthetic templates and it can fit redshifts using a wider range of multiwavelength data. Their work has the advantage of having been applied to larger fields. However, high signal-to-noise data may not be available in all bands for these large areas and thus, their resulting redshifts have a typical accuracy of about 10\%, significantly worse than those obtained here. Our approach  depends on high-quality and homogeneous input data, only available for very limited areas of the sky. By combining the results of both works it is possible to:  1) determine overall trends and the effects of cosmic variance based on the approach of P\'erez-Gonz\'alez et al.(2005); and 2) prove the redshift distribution more accurately in a more limited sky region based on the approach in this paper.

\section{The redshift distribution of $\rm 24 \mu \lowercase{m}$ galaxies in the GOODS/CDFS} 
\label{sec_redsh} 
\begin{figure} 
\plotone{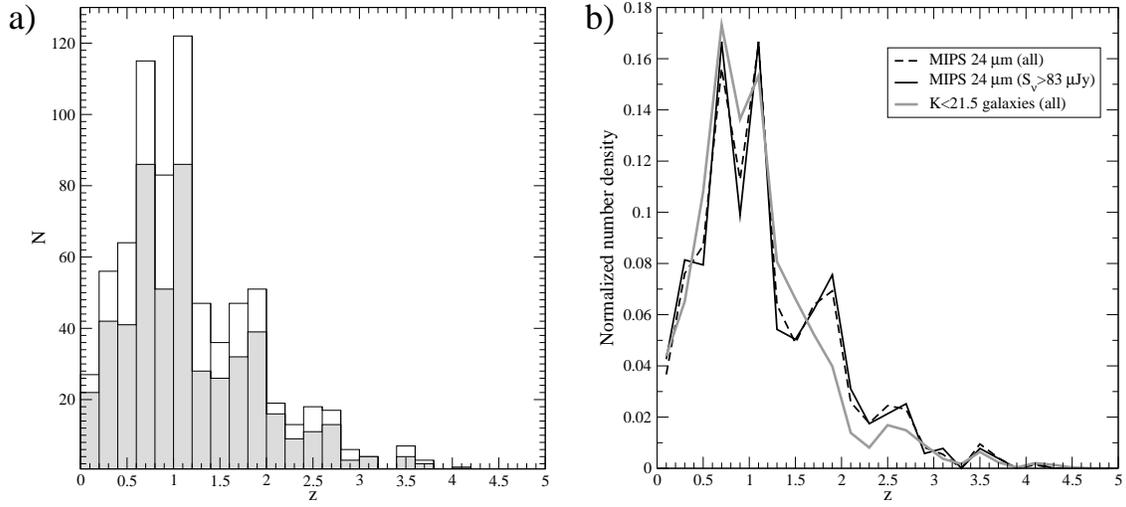} 
\caption{ a) The redshift distribution of the MIPS  $\rm 24\, \mu m $ galaxies with $K_s<21.5$ counterparts in the GOODS/CDFS: all sources and sources with $S_\nu >83 \, \mu {\rm Jy}$ (empty and shaded histograms, respectively). b) The normalized redshift distributions of all and $S_\nu >83 \, \mu {\rm Jy}$  $\rm 24\, \mu m $ galaxies (black dashed and solid lines, respectively), compared to the normalized redshift distribution of the total $K_s<21.5$ sample in the same field (grey solid line).  
\label{24zhisto}} 
\end{figure}

Figure~\ref{24zhisto}a) shows the redshift distribution of the MIPS $\rm 24\,\mu m$ galaxies with $K_s<21.5$ counterparts in the GOODS/CDFS. The empty (shaded) histogram corresponds to all ($S_\nu > 83 \, \mu {\rm Jy}$) sources.  The peaks in the redshift distribution at $z \sim 0.7$ and $z \sim 1.1$ are due to the effect of large-scale structure in the CDFS (e.g. Le F\`evre et al. 2004).  We find that 72\% of the $\rm 24\,\mu m$ galaxies lie at redshifts $z<1.5$, while the remaining 28\% are found to be at $z \geq 1.5$ (the percentages are similar for the samples containing all and only $S_\nu > 83 \, \mu {\rm Jy}$ galaxies).  This confirms the existence of a substantial population of mid-IR sources at high redshifts. Approximately half of the $z \geq 1.5$ sources have the characteristic colours of Extremely Red Galaxies (ERGs), $(I_{775} - K_s) >4.0$ (Vega), indicating that an important fraction of the high-redshift mid-IR background is constituted by optically obscured sources, as possibly expected (cf. Yan et al.~2004).

 Figure~\ref{24zhisto}b) shows the normalized redshift distributions of the MIPS $\rm 24\,\mu m$ galaxies with $K_s<21.5$ counterparts (black dashed and solid lines for all and $S_\nu > 83 \, \mu {\rm Jy}$ sources, respectively), compared to the redshift distribution of the total $K_s<21.5$ galaxy population (grey solid line) in the same field. Several features are present in all the three curves, which are the consequence of cosmic variance effects. In contrast, we observe the presence of a depression in the redshift distribution  of $\rm 24\,\mu m$ galaxies at redshift $z \sim 1.5$ and a bump at redshift $z \sim 1.9$, both of which do not appear in the total $K_s<21.5$ galaxy curve.

\begin{figure} 
\plotone{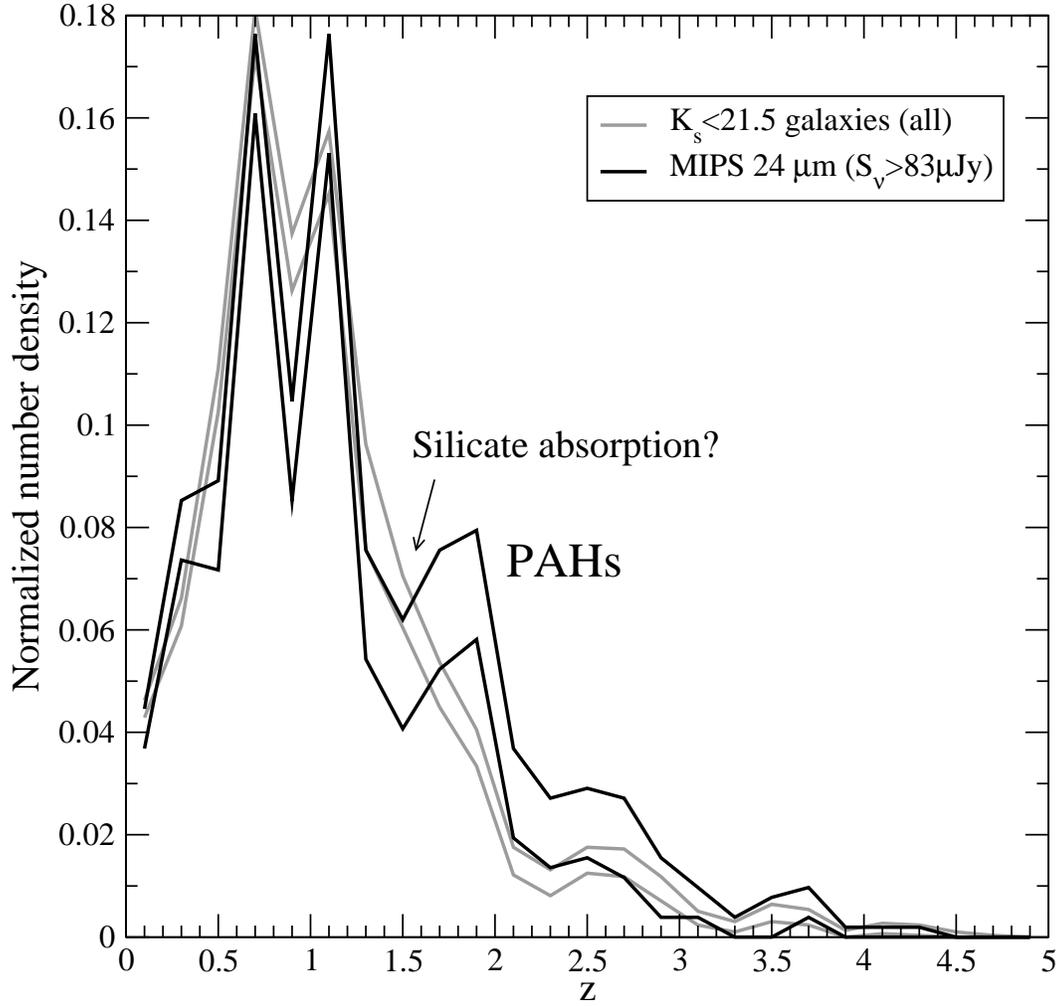} 
\caption{  The 95\% confidence limits on the normalized redshift distributions of the MIPS $S_\nu(24 \, \mu \rm m) >83 \, \mu {\rm Jy}$   galaxies (black lines) and  all the $K_s<21.5$ galaxies (grey lines). The bump in the redshift distribution of $\rm 24\, \mu m$ galaxies at $z \sim 1.9$ indicates the presence of PAH emission at these high redshifts. The marginally significant depression at redshift $z \sim 1.5$ might be due to silicate absorption.
\label{signifzdist}} 
\end{figure}

  To assess the significance of the features observed exclusively on the $\rm 24\,\mu m$ galaxy redshift distributions, we computed confidence limits on the different curves. We performed Monte-Carlo simulations to create 1000 mocked redshift catalogs, alternatively for the total $K_s<21.5$ galaxy population and for the MIPS $\rm 24\,\mu m$  galaxies with flux $S_\nu > 83 \, \mu {\rm Jy}$. We constructed each mock catalog assigning to each source a random redshift, with a gaussian probability centred at the original redshift $z$ of the source and a dispersion equal to $0.02 \, (1+z)$ (i.e. the median error at the corresponding redshift). The redshifts of those sources with spectroscopic values were left fixed. We re-computed the normalized redshift distribution for each mock catalog and determined confidence limits on the original distribution curves. The 95\% confidence limits on the total $K_s<21.5$ galaxy and the MIPS ($S_\nu > 83 \, \mu {\rm Jy}$) galaxy distributions are shown in Figure~\ref{signifzdist}.

  Figure~\ref{signifzdist} shows that, even taking into account the error bars, the peak in the  $\rm 24\,\mu m$ redshift distribution at $z \sim 1.9$ is significant, lying $\sim 4 \sigma$ from the original  total $K_s<21.5$ galaxy curve. Thus, we conclude that the redshift distribution of $\rm 24\,\mu m$ galaxies presents a  real secondary bump at these high redshifts. This secondary peak has been predicted by Lagache et al.~(2004) and is the consequence of the selection effect produced by the  presence of  PAH emission features entering the observed $\rm 24\,\mu m$ band.  Given the width of the $\rm 24\,\mu m$ filter (whose transmission covers the wavelength range $\sim 20-28 \, \mu \rm m$), both the $7.7 \mu \rm m$ and the $8.6 \mu \rm m$ PAH lines could contribute to the redshift distribution peak observed at $z \sim 1.9$.  Our results allow us to conclude that PAH molecules must be already present in star-forming galaxies at high redshifts.

  The confidence limits shown in Figure~\ref{signifzdist} also indicate that the depression observed in the  $\rm 24\,\mu m$ redshift distribution at $z \sim 1.5$ is only marginally significant within our sample. The existence of such a depression could be interpreted as due to the $\rm 9.8 \, \mu m$ silicate absorption feature  entering the  $\rm 24\,\mu m$ filter at this redshift, which has been observed for some galaxies in previous works (Houck et al.~2005; Yan et al.~2005). However, given the errors in the redshift distributions presented here, we cannot reach any firm conclusion on the possible selection effect produced by silicate absorption at high redshifts.

  We also see a substantial deficit of $\rm 24\,\mu m$ sources at redshift $z \sim 0.9$, with respect to the total $K_s<21.5$ sources.  At this redshift, a positive selection effect on $\rm 24\,\mu m$ galaxies would be expected due to the presence of the $\rm 12.7 \, \mu m$ PAH emission line. The observed relative deficit of bright mid-IR selected galaxies indicates that, within our sample, star-forming galaxies are present in a minor proportion at this redshift. This is quite likely due to a mere cosmic variance effect. However, we note that this deficit in our $\rm 24\,\mu m$ sample occurs at the same redshift where there is a dip in the total $K_s<21.5$ redshift distribution. If we consider that galaxy interactions are a triggering mechanism for star-formation, then it will be possible  that star-formation activity is particularly inhibited  in regions with an under-density of objects. The study of similar regions in other areas of the sky are necessary to determine whether the two facts are actually related.

\begin{figure} 
\plotone{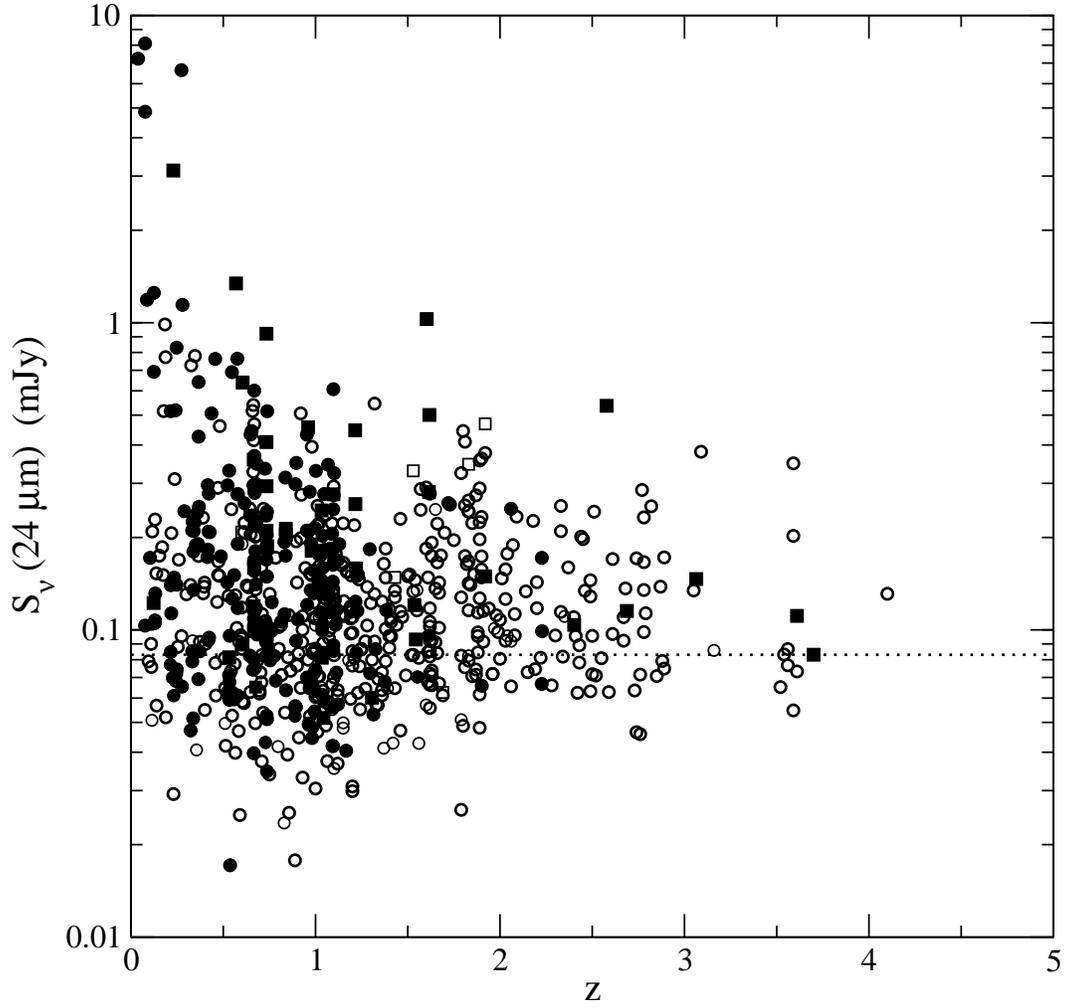} 
\caption{ $\rm 24\, \mu m $ flux versus redshift for the MIPS $\rm 24\, \mu m $ galaxies in the GOODS/CDFS associated to a $K_s<21.5$  counterpart. The symbols are the same as in figure~\ref{24vsKs}. Filled symbols indicate spectroscopic redshifts.
\label{24vsz}} 
\end{figure}

  Evidence of the existence of PAH emission in the spectra of a few high-redshift $1.7-1.8 < z < 2.6-2.8$ galaxies has recently been presented by Houck et al.~(2005), Yan et al.~(2005) and Lutz et al.~(2005). In this work, we extend this evidence through the study of the redshift distribution of the whole $\rm 24 \, \mu m$ galaxy population.  It is interesting to note, however,  that both Houck et al.~(2005) and Yan et al.~(2005) found that only a minority of their galaxies at $z \sim 1.7-2.8$ showed clear PAH emission features in their spectra, while most of their remaining high-redshift galaxies were  AGN-dominated or had silicate absorption features. The difference in the nature of the sources dominating Houck et al. and Yan et al. samples can be explained taking into account that their objects are on average an order of magnitude more luminous than most of the sources analyzed in this work. Deep  $\rm 24\,\mu m$ samples are necessary to uncover a substantial population of PAH galaxies at redshift $z \sim 2$. Figure~\ref{24vsz} shows the $\rm 24\,\mu m$ flux of each galaxy $S_\nu$ versus redshift $z$. The symbols are the same as in Figure~\ref{24vsKs}. Filled symbols correspond to sources with spectroscopic redshifts.  Within our sample, we see that only a few bright $S_\nu(24 \, \mu m)>0.5 \, {\rm mJy}$ sources are placed at  high redshift ($z \gsim 1.5$) and they are all active galaxies. At fainter fluxes $S_\nu \lsim 0.3 \, {\rm mJy}$, starbursts  produce the bulk of the  mid-IR emission at high redshifts.

   P\'erez-Gonz\'alez et al.~(2005) found that 24\% of the $\rm 24\,\mu m$ galaxies with  $S_\nu > 83 \, \mu {\rm Jy}$ in an extended region of the CDFS and the Hubble Deep Field North (HDFN) were at redshifts $z>1.5$, in agreement with our value of 28\% within the errors and taking into account the different fields surveyed. However, their redshift distribution does not show the PAH-induced peak at $1.6 \lsim z \lsim 2.2$. They suggested that this feature may have been blurred by the errors in photometric redshifts; this suggestion is confirmed by our detection of the PAH bump using redshifts of higher accuracy. Taken together, the two studies show that the drop in the number of sources beyond $z \sim 1.2$ holds generally and is not just observed in the CDFS. In addition, we show that there is a significant peak produced by PAH emission on top of this general trend.

\begin{figure} 
\plotone{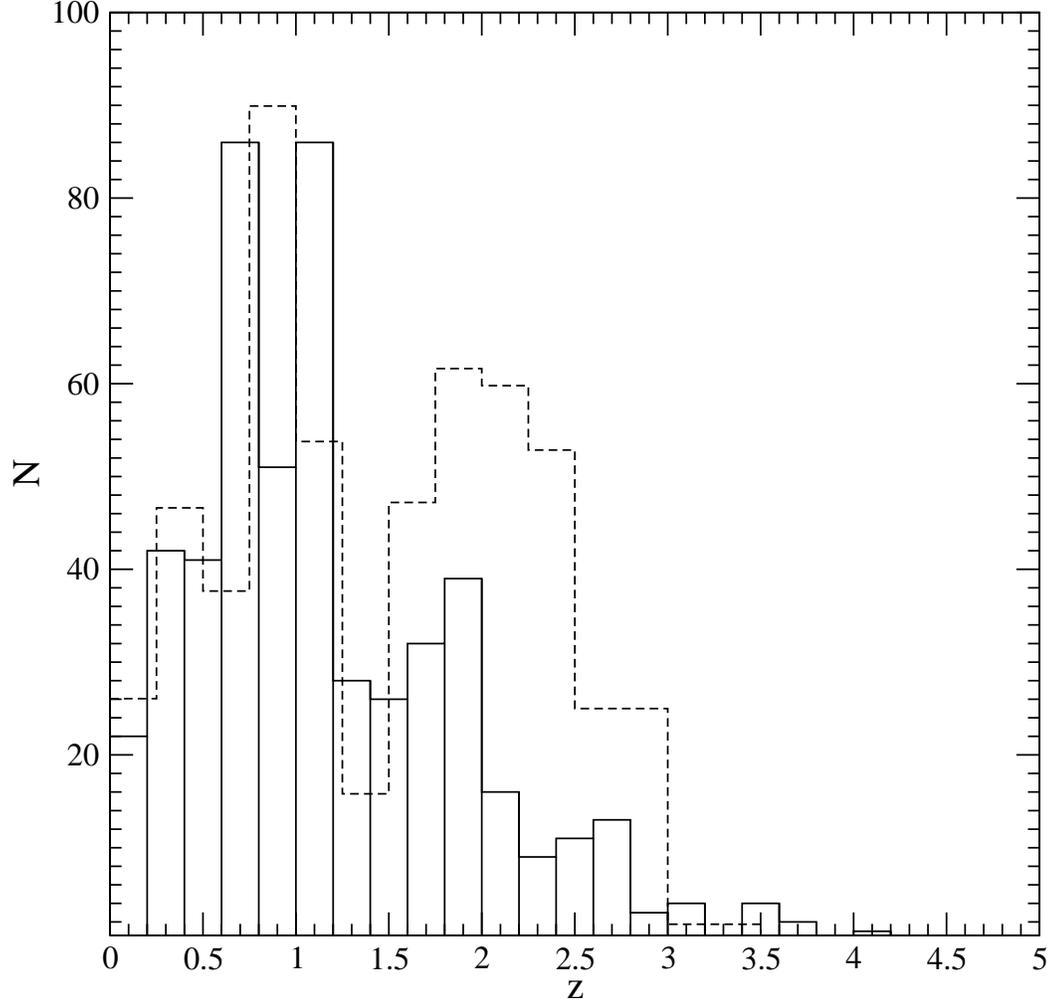} 
\caption{ The comparison of the observed redshift distribution of MIPS $\rm 24\, \mu m $ galaxies in the GOODS/CDFS (solid histogram) with the redshift distribution predicted by Lagache et al.(2004) (dashed histrogram). Both distributions correspond to sources with flux $S_\nu(24 \, \mu {\rm m}) >83 \, \mu {\rm Jy}$ in an area of 131 arcmin$^2$.
\label{guilmod}} 
\end{figure}

  Figure~\ref{guilmod} shows the comparison of our observed redshift distribution of $\rm 24\,\mu m$ galaxies in the GOODS/CDFS (solid histogram) with the distribution predicted by  Lagache et al.(2004) using the Lagache, Dole \& Puget (2003) model (dashed histogram). Both distributions correspond to sources with  flux $S_\nu > 83 \, \mu {\rm Jy}$ in an area of 131 arcmin$^2$. The observed and predicted distribution have 521 and 518 galaxies, respectively, confirming that the Lagache et al.~(2003) model very well reproduce the observed number counts.  Also in agreement with the model, we find that a significant population of $\rm 24\,\mu m$ galaxies lie at redshifts $z>1.5$, and our accurate redshift determinations reveal the  predicted secondary peak in the redshift distribution at $z \sim 1.9$  produced by the presence of PAH emission. However, the fraction of galaxies at redshifts  $z>1.5$ is significantly overpredicted by the Lagache et al. model. Other models (e.g. Chary \& Elbaz 2001) basically predict a negligible fraction of sources at redshifts $z>1.5$ and also do not fit the observed redshift distribution. Therefore, theoretical models require further refinements to reproduce the distant infrared-emitting galaxy populations.

\section{The assembled stellar masses of MIPS $\rm 24 \mu \lowercase{m}$ galaxies} 
\label{sec_stmass}
  
\begin{figure} 
\plotone{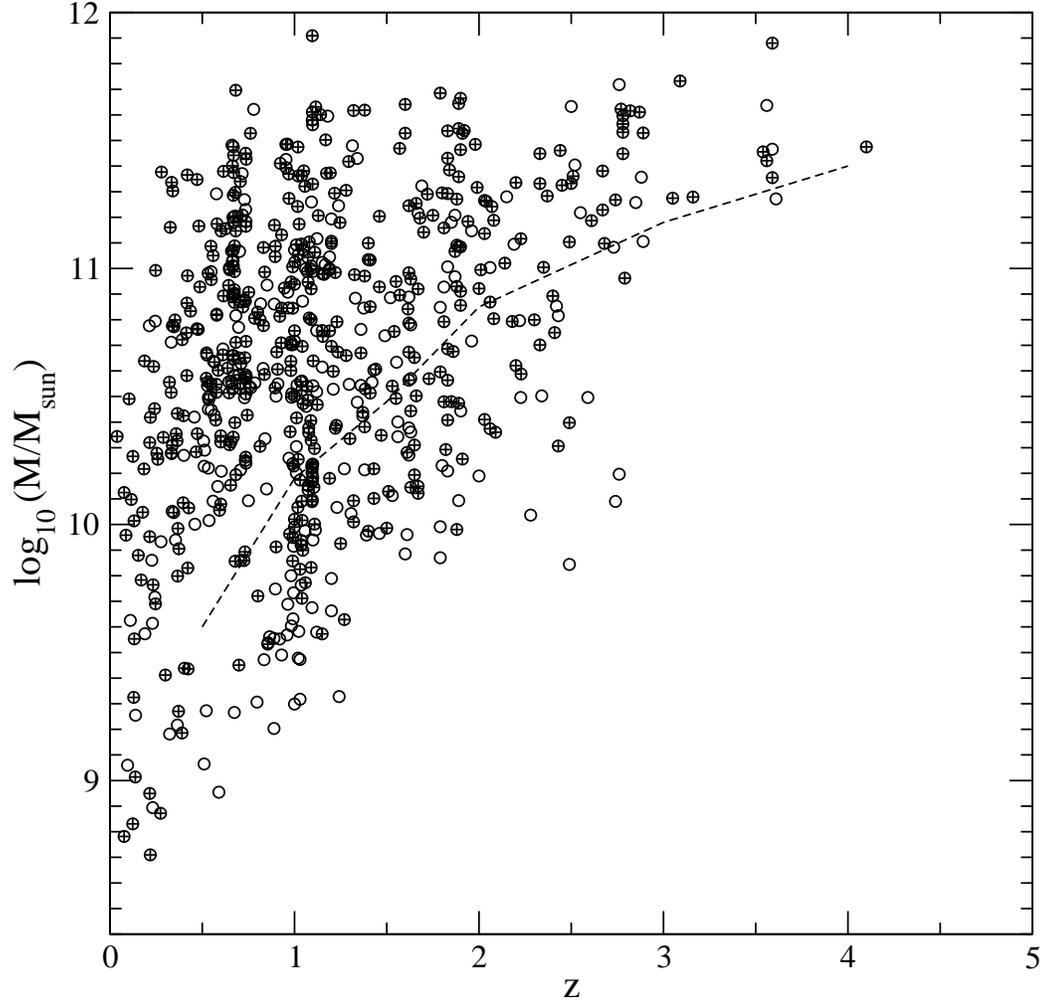} 
\caption{ The estimated assembled stellar masses for the MIPS $\rm 24\, \mu m $ galaxies with $K_s<21.5$ counterparts in the GOODS/CDFS. The symbols with a cross correspond to galaxies with flux $S_\nu(24 \, \mu m) >83 \, \mu {\rm Jy}$. The dashed line indicates the estimated mass completeness limits imposed by the $K_s<21.5$ survey. However, we note that these limits are basically irrelevant for the galaxies with flux $S_\nu(24 \, \mu m) >83 \, \mu {\rm Jy}$, as the $K_s$-band  identification completeness for these galaxies is $\sim$94\%. 
\label{stmassvsz}} 
\end{figure}

    Figure~\ref{stmassvsz} shows the estimated assembled stellar masses of the MIPS $\rm 24 \, \mu m$ galaxies versus redshift $z$ in the GOODS/CDFS. The estimation of the stellar masses is based on the optical-to-near-IR SED fit of each galaxy at the determined  redshift (either spectroscopic, from COMBO17 or HYPERZ/BPZ) and is completely independent of its $\rm 24 \, \mu m$ properties.  We determined estimated stellar masses for all the $\rm 24 \, \mu m$ galaxies with a $K_s<21.5$ identification, except for any X-ray-classified AGN/QSO or any other X-ray source without a suitable GISSEL98 template  optical-to-near-IR SED fit. We excluded the latter objects to avoid galaxies with partial  contamination of the optical-to-near-IR light by a hidden AGN.    The  stellar masses are computed from the modelled  rest-frame $K_s$-band galaxy luminosity of each galaxy,  where the mass-to-light ratios have a minimum dependence on the SED star-formation histories or dust corrections (see Caputi et al. 2005a,b for further details). The resulting stellar mass estimates are typically accurate within a factor $\lsim 2$.   A single power-law Salpeter IMF over stellar masses $M=(0.1-100) \, M_\odot$ has been assumed throughout. The estimated mass-completeness limits for counterparts of the MIPS $\rm 24 \, \mu m$ galaxies are $1.5 \times 10^{10} M_\odot$,   $7.0 \times 10^{10} M_\odot$ and $1.5 \times 10^{11} M_\odot$ at redshifts $z=1,2,3$, respectively, based on the $K_s=21.5$ magnitude limit and the median of the k-corrections. However, it should be noted that the mass completeness limits are basically irrelevant for the sources with flux $S_\nu > 83 \, \mu {\rm Jy}$ (symbols filled with a cross in Figure~\ref{stmassvsz}), as the $K_s$-band identifications are almost complete above this limit.

     Inspection of figure \ref{stmassvsz} shows that, at least at redshifts $z \lsim 1$,  star-formation activity  takes place in galaxies of a wide range of assembled stellar masses, from $\sim 10^9 \, M_\odot$ to $\sim 10^{12} \, M_\odot$. At high redshifts $z \gsim 2.5$, the existence of a significant population of massive $M>10^{11} M_\odot$  star-forming galaxies is revealed in the mid-IR at the depth of the GTO/CDFS images. It is now commonly believed that a fraction of present-day massive galaxies is already in place at high redshifts (Cimatti et al. 2004; Glazebrook et al. 2004; Caputi et al. 2005a,b).  However, it is still unclear when the bulk of the stars in these massive galaxies was formed (cf. Papovich et al. 2005). The identification of  massive objects on the MIPS $\rm 24 \mu m$ images can put constraints on their star-formation histories and the amount of stellar mass built-up at different redshifts. 
%
\section{Constraints on the star-formation history} 
\label{sec_stform}

\subsection{The evolution of star-forming galaxies with redshift}

  The study of the physical properties of IR-selected galaxies at different redshifts is of importance to understand how star-formation evolved with cosmic time. In this section, we use the $\rm 24 \, \mu m$  fluxes of the  galaxies in the GOODS/CDFS in conjunction with the redshift and stellar mass estimations, to study the evolution of  star-formation activity.

  Chary \& Elbaz (2001) and Elbaz et al. (2002) showed that the mid-IR luminosities of nearby galaxies were correlated (with some scatter) with their bolometric IR luminosity $L_{IR}$,   defined as $L_{IR}=L(8-1000 \mu{\rm m})$. They fitted the following set of relations:
   
\begin{eqnarray}
\label{chelb}
L_{IR}&=&11.1^{+5.5}_{-3.7} \times (\nu L_\nu[15 \mu m])^{0.998}, \\
&=&0.89^{+0.38}_{-0.27} \times (\nu L_\nu[12 \mu m])^{1.094}, \\
&=&4.78^{+2.37}_{-1.59} \times (\nu L_\nu[6.75 \mu m])^{0.998} \\
&& {\rm for} \,\,  \nu L_\nu[6.75 \mu m]< 5 \times 10^9 L_\odot, \nonumber \\
&=&4.37^{+2.35}_{-2.13} \times 10^{-6} \times (\nu L_\nu[6.75 \mu m])^{1.62} \\
&& {\rm for} \,\,  \nu L_\nu[6.75 \mu m]\geq  5 \times 10^9 L_\odot, \nonumber
\end{eqnarray}    
   
\noindent where all the luminosities are in solar units. There are many indications that the overall spectral energy 
distributions of high-redshift infrared galaxies are similar to the local ones captured in  
equations (1) to (4). For example, the local far-IR/radio and mid-IR/radio correlations 
(Condon 1992) still hold at higher redshifts (Appleton et al. 2004). Composite SEDs of 
high redshift infrared galaxies resemble local templates (e.g. Egami et al. 2004). The MIPS observations in the CDFS indicate that, at  least to $z\sim 1$, galaxies have 24 and 70~\micron\  flux ratios that follow  the local distribution as a function of total IR luminosity. In addition, number-count models assuming similar behaviour can fit the 24, 70, and 160$\mu$m data simultaneously (Lagache et al. 2004).

\begin{figure} 
\plotone{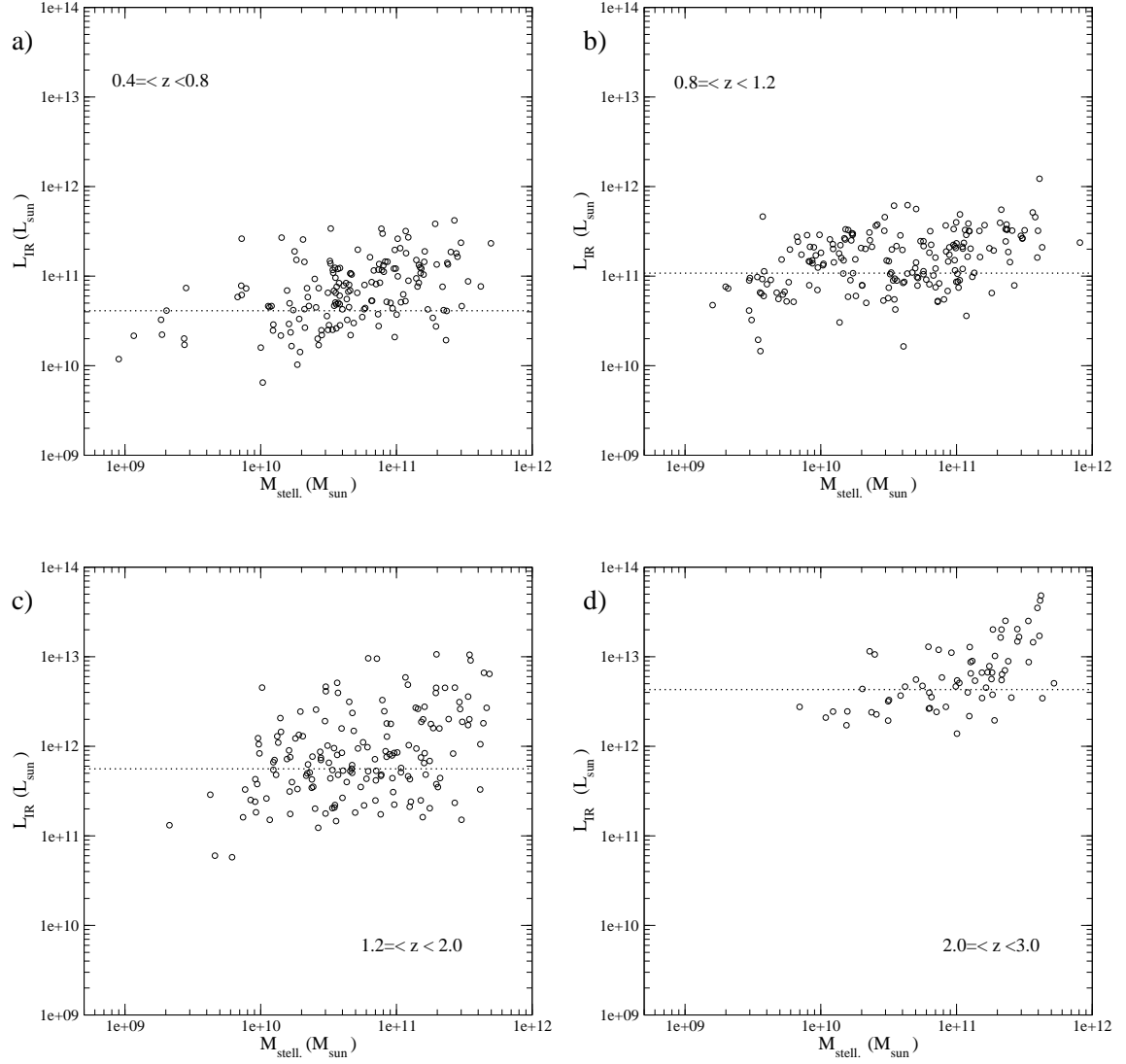} 
\caption{ Bolometric infrared luminosities versus assembled stellar masses for the MIPS $\rm 24\, \mu m $ galaxies with $K_s<21.5$ counterparts in the GOODS/CDFS, in different redshift bins: a) $0.4 \leq z < 0.8$; b) $0.8 \leq z < 1.2$; c) $1.2 \leq z < 2.0$; d) $2.0 \leq z < 3.0$. The dotted lines delimit the region of  completeness at the mean redshift of each bin, taking into account  the $S_\nu(24 \, \mu m)=83 \mu {\rm Jy}$ limit and the fact that our $K_s$-band identifications are almost complete above that flux (i.e. the mass completeness limits imposed by the $K_s=21.5$ cut are basically irrelevant above that limit).
\label{LM_comp}} 
\end{figure}

 We used the $\rm 24 \, \mu m$ fluxes of the MIPS galaxies with a $K_s<21.5$ counterpart in the GOODS/CDFS to compute their bolometric luminosities $L_{IR}$, assuming the Chary \& Elbaz  relations in eqs. (1) to (4). We excluded all the known AGNs/QSOs from the present analysis.  We considered    that the $\rm 24 \, \mu m$ flux mapped the rest-frame $\rm 15 \mu m$, $\rm 12 \mu m$  and $\rm 6.75 \mu m$ fluxes in the redshift ranges $0.4 \leq z <0.8$,  $0.8 \leq z <1.2$ and $2.0 \leq z <3.0$, respectively. In the redshift range $1.2 \leq z <2.0$, we assumed that the  infrared luminosity $L_{IR}$ was given by the average of the luminosities obtained using eqs. (2) and (3-4).
 Figure~\ref{LM_comp} shows the bolometric IR luminosities of the  $\rm 24 \, \mu m$  galaxies versus assembled stellar mass in  different redshift bins. The dotted lines delimit the region of  luminosity completeness at the mean redshift of each bin, taking into account  the $S_\nu(\rm 24 \, \mu m)=83 \, \mu {\rm Jy}$ limit and the fact that our $K_s$-band identifications are almost complete above that flux (i.e. the mass completeness limits imposed by the $K_s=21.5$ cut are basically irrelevant above that limit).

 Although for clarity the error bars are not shown in Figure~\ref{LM_comp}, we estimated the errors on our computed bolometric luminosities using an independent set of models. We integrated the different Dale et al.~(2001) and Dale \& Helou~(2002) SED templates, normalized to our observed $24 \, \rm \mu m$ fluxes, to obtain the corresponding IR luminosities as a function of  redshift. The complete set of the Dale et al. SEDs consists of 64 templates, each one characterized by a parameter $\alpha$ ($0.0625<\alpha<4.0$), which depends on the combination of  the observed local SEDs used to construct these models. Compared to the values obtained with the Dale et al. templates,  our luminosities based on the Chary \& Elbaz formulae appear to be accurate within a factor 2-3  to redshift $z \approx 1.5$. Above that redshift, the consideration of the complete set of the Dale et al. SEDs would yield uncertainties of up to a factor 10-20 (Dale et al. 2005).  However, Dale et al. ~(2005) recently used the Dale et al. templates to fit the observed 1-850 $\rm \mu m$  photometry of a sample of 75 local galaxies. The majority of these galaxies ($\gsim 75$\%) were best-fitted by templates with $2.0<\alpha<4.0$. As we explained above, there are many indications that the overall spectral energy distributions of high-redshift IR galaxies are similar to the local ones. Thus, this subset of the Dale et al. templates should be representative of the majority of the IR galaxy types  also at high redshifts. If we restrict the comparison of our luminosities computed with the Chary \& Elbaz formulae to those values obtained with the Dale et al. $2.0<\alpha<4.0$ templates, we find that the error bars would still be within a factor 2-3 up to redshift $z \approx 2$, and a factor 5 would be sufficient to account for the errors above that redshift. These factors should be representative of the error bars on the IR luminosities of the majority of our galaxies, including the errors introduced by the k-corrections within each of the considered redshift bins.

 We observe that the evolution of the IR luminosities with redshift is still very significant even taking into account the error bars.  Panel a) shows that most of the  $\rm 24 \, \mu m$ galaxies at redshifts $0.4 \leq z < 0.8$ have infrared luminosities $L_{IR}<10^{11}\, L_{\odot}$.  The maximum observed infrared luminosities increase with redshift, and luminous infrared galaxies (LIRGs) characterized by  $10^{11}\, L_{\odot}<L_{IR}<10^{12}\, L_{\odot}$  \cite[]{sand96} are the dominant  $\rm 24 \, \mu m$  sources at redshifts  $0.8 \leq z < 1.2$ at the depth of the CDFS MIPS/GTO images (Le Floc'h et al. 2005). The majority of the mid-IR sources at  $0.4 \leq z < 1.2$ are hosted by intermediate-mass galaxies with stellar masses $10^{10} \, M_{\odot} \lsim M \lsim 10^{11}  \, M_{\odot}$, in agreement with recent findings by Hammer et al.(2005), although some more massive galaxies could also be classified as LIRGs at these redshifts. Within our surveyed area, there is virtually no ultra-luminous infrared galaxy (ULIRG) with $L_{IR}>10^{12}\, L_{\odot}$ at $z<1.2$. ULIRGs might be present at these low redshifts, but are indeed very rare (e.g. Flores et al.~1999).  At $1.2 \leq z \leq 2.0$, ULIRGs start to be a very significant population ($\sim$ 65\% at $S_\nu > 83 \mu {\rm Jy}$) and they are also hosted by intermediate to high-mass galaxies. At $z>2$, the limits of  the  $\rm 24 \, \mu m$  survey only allow us to explore the bright IR-luminosity  end of the star-formation activity. We observe sources with extremely high infrared luminosities  $10^{12}\, L_{\odot}<L_{IR}<10^{14}\, L_{\odot}$ mainly harboured by galaxies with stellar masses  $M > 10^{11} \, M_{\odot}$.  Our results on luminosity and SFR evolution agree well with the studies for larger fields made by Le Floc'h et al.~(2005) and P\'erez-Gonz\'alez et al.~(2005).

  Thus,  figure~\ref{LM_comp} confirms the global trend of a high degree of evolution in the star-formation activity from low to high redshifts.  The extremely high IR luminosities observed at $z \gsim 2$ show that star-formation has been a much more violent process in the past and these violent episodes of star-formation must have built up a significant part of the stellar mass content of at least a fraction of present-day massive galaxies. We explore in more detail this possibility in the next section, where we analyze and constrain the typical timescales for star-formation of galaxies with different assembled stellar masses at different redshifts.

\subsection{Probing star-formation in different-mass galaxies}

 An estimator of the instantaneous star-formation rate (SFR) of a galaxy based on its bolometric infrared luminosity has been obtained by Kennicutt (1998):

\begin{equation}
\label{kennsfr}
SFR=1.72 \times 10^{-10} \, L_{IR},
\end{equation}

\noindent  where the star-formation rate $SFR$ is given in $M_\odot/{\rm yr}$, the luminosity $L_{IR}$ is in solar units and  a Salpeter IMF over stellar masses $M=(0.1-100) \, M_\odot$ is assumed.  We used equation (\ref{kennsfr}) to obtain an estimate of the instantaneous SFR  in the  $\rm 24 \, \mu m$  galaxies studied here, and to explore the relation of this  SFR  with the already assembled stellar mass of each galaxy, at different redshifts.
 
\begin{figure} 
\plotone{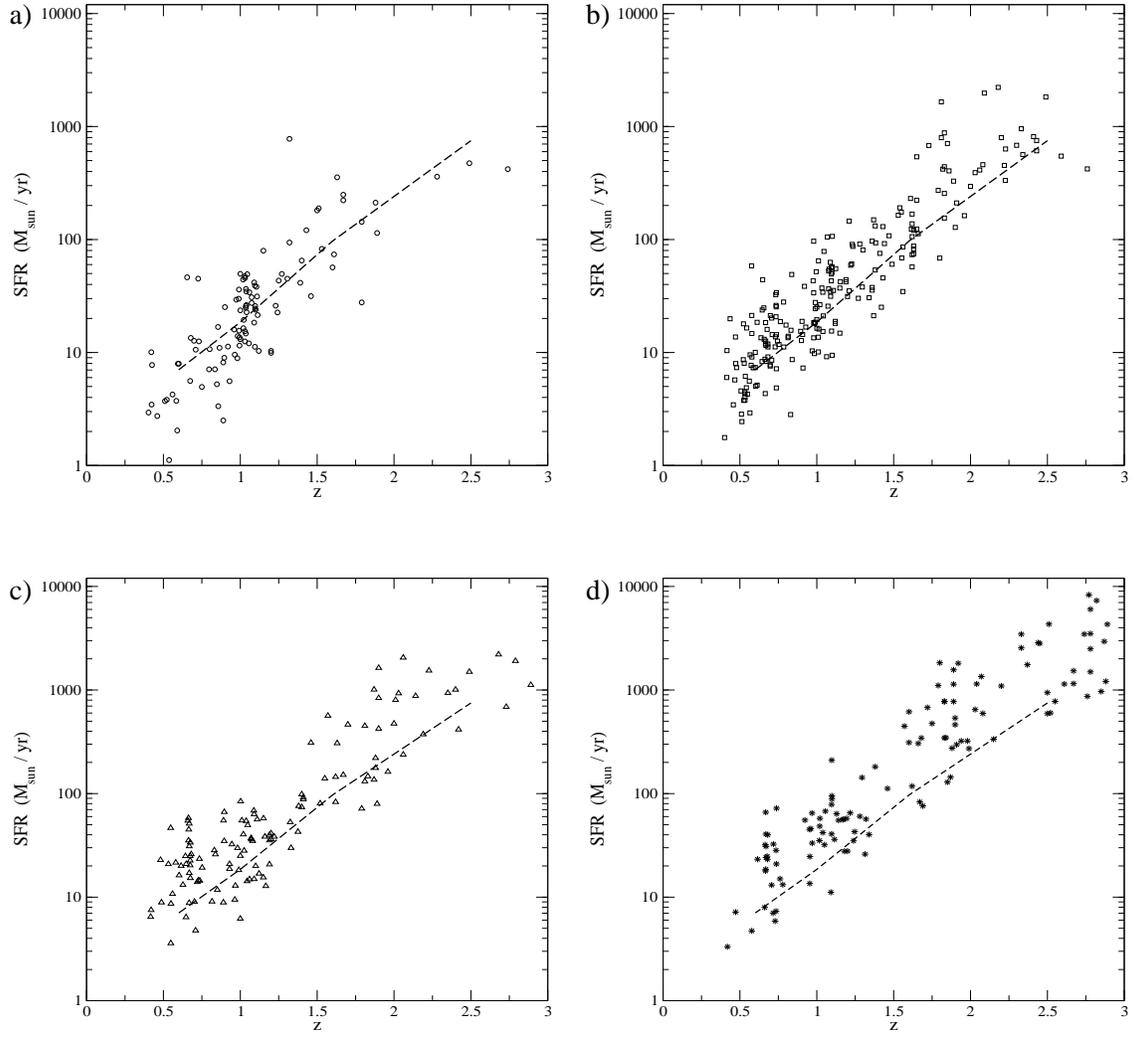} 
\caption{ The evolution of the star-formation rate of galaxies with different assembled stellar mass: a) $M<1.5 \times 10^{10} M_\odot$; b) $1.5 \times 10^{10} M_\odot \leq M < 7.0 \times 10^{10} M_\odot$; c) $7.0 \times 10^{10} M_\odot \leq M < 1.5 \times 10^{11} M_\odot$; d) $M \geq 1.5 \times 10^{11} M_\odot$. In all the panels, the dashed line delimits the region of completeness of the observed star-formation rates as imposed by the flux limit $S_\nu(24 \, \mu m)=83 \mu {\rm Jy}$.
\label{SFRvsz_mass}} 
\end{figure}

  Figure~\ref{SFRvsz_mass} shows the evolution of the SFR with redshift for galaxies with assembled stellar masses in different mass bins. In all the panels, the dashed line delimits the region of completeness of the observed SFR versus redshift, as imposed by the flux limit $S_\nu(24 \, \mu {\rm m})=83 \mu {\rm Jy}$. We find that the IR-derived SFR exponentially grows with redshift, independently of the galaxy stellar mass (of course the lower envelope of SFR is a consequence of the limits of the $\rm 24 \, \mu m$ survey, but the maximum observed values do exponentially grow with redshift).  However, panel a) shows that, at a given redshift,  most of the lowest mass galaxies have relatively small SFR, which are at most marginally above the completeness limit of the $\rm 24 \, \mu m$ survey analyzed here.  This effect is particularly evident at high redshifts $z \gsim 2$,  where we observe  a lack of galaxies with stellar mass $M<1.5 \times 10^{10} M_\odot$ and SFR above the completeness limit. Unless the minor fraction of unidentified sources with $S_\nu(24 \, \mu {\rm m})>83 \mu {\rm Jy}$ constitutes a biased population of low mass galaxies at high redshifts,  we conclude that low-mass objects with very high SFR ($SFR \gsim 500 \, M_\odot/{\rm yr}$) are indeed rare at $z \gsim 2$.  On the contrary,  galaxies with greater stellar mass $M \geq 1.5 \times 10^{10} M_\odot$  and $SFR \gsim 500 \, M_\odot/{\rm yr}$  are progressively found at higher redshifts. 
  
\begin{figure} 
\plotone{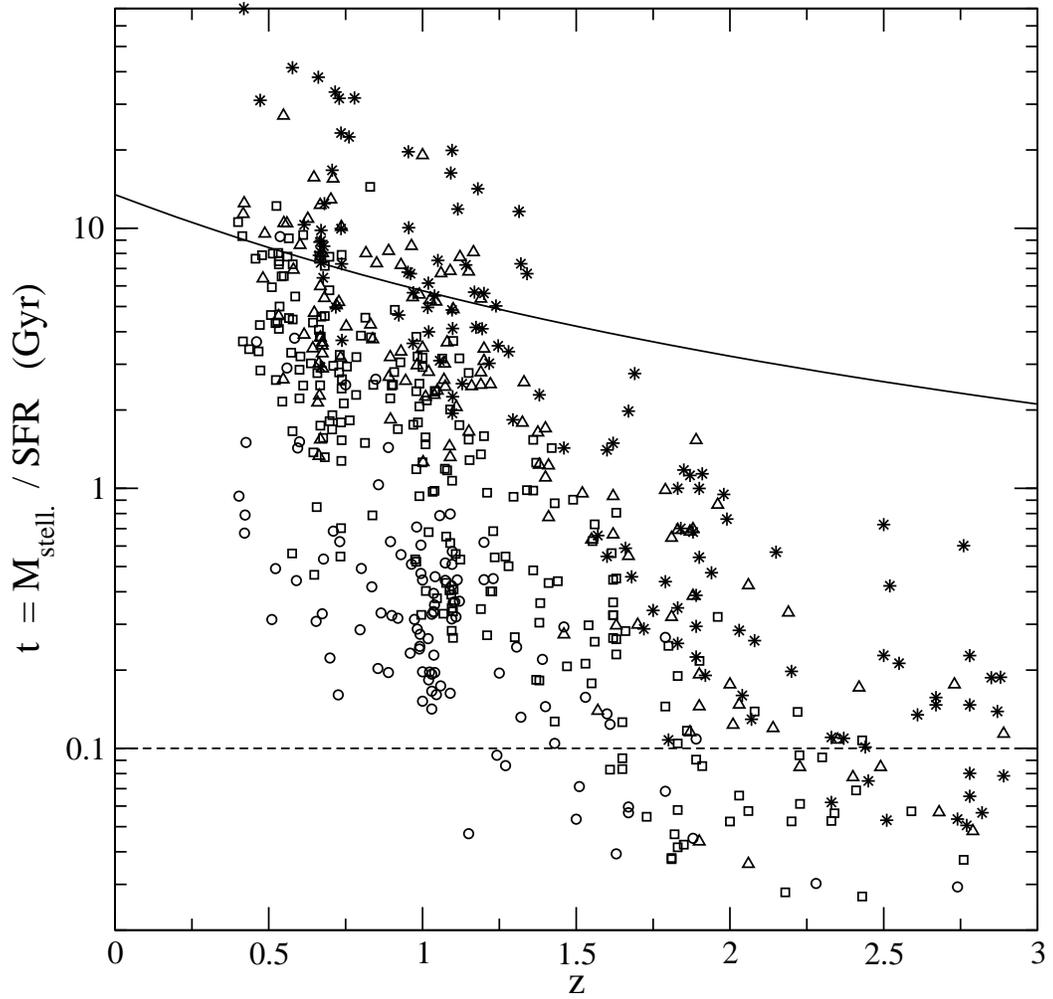} 
\caption{ The ratio between the assembled stellar mass and the instantaneous star-formation rate of  MIPS $\rm 24\, \mu m $ galaxies with $K_s<21.5$ counterparts in the GOODS/CDFS, versus redshift. Different symbols indicate different assembled stellar mass ranges: $M<1.5 \times 10^{10} M_\odot$ (circles),  $1.5 \times 10^{10} M_\odot \leq M < 7.0 \times 10^{10} M_\odot$ (squares), $7.0 \times 10^{10} M_\odot \leq M < 1.5 \times 10^{11} M_\odot$ (triangles) and $M \geq 1.5 \times 10^{11} M_\odot$ (asterisks). The solid line shows the age of the Universe as a function of redshift, while the dashed line indicates an upper limit to the typical lifetime of a starburst ($\rm 0.1 \, Gyr$).
\label{stmsfrvsz}} 
\end{figure}

  Figure~\ref{stmsfrvsz}  shows the ratio of already assembled stellar mass over the instantaneous SFR, $M/SFR$,  versus redshift $z$ for the  $\rm 24 \, \mu m$ galaxies with a $K_s<21.5$ counterpart in the GOODS/CDFS.    Different symbols correspond to galaxies with stellar masses in different mass ranges, as in Figure~\ref{SFRvsz_mass}.  The solid line shows the age of the Universe as a function of redshift, while the dashed line indicates the upper limit of the typical lifetime of a starburst ($\rm \sim 0.1 \, Gyr$). The assembled stellar masses have been computed as in Section~\ref{sec_stmass} and are completely independent of the $\rm 24 \, \mu m$ properties of each galaxy.  If the SFR in a galaxy were larger in the past or at most stayed constant over cosmic time,  the ratio $M/SFR$ would give an upper limit for the lifetime of the source.   Therefore, figure~\ref{stmsfrvsz} allows  putting constraints on the star-formation history of the MIPS  $\rm 24 \, \mu m$ galaxies present at different redshifts.

  A first striking conclusion is that, at high redshifts $z \sim 2-3$,  a starburst lifetime appears to be sufficient to construct a substantial fraction and up to the whole amount of the stellar mass of some massive galaxies. If equations (1) to (5) are still applicable at those redshifts, very high SFR are predicted, ranging from $\sim 500 M_\odot/{\rm yr}$ to $5000 M_\odot/{\rm yr}$. In this way, a stellar mass of $\approx 10^{10} M_{\odot} - 10^{11} M_{\odot}$ could be constructed in a  $\rm \sim 0.01 \, Gyr - 0.1 \, Gyr$ period of time. However, this does not necessarily mean that all these systems are being formed at $z \sim 2-3$.  Recent works suggest that a non-negligible fraction of present-day massive galaxies would be present before even higher ($z \gsim 3$) redshifts (van Dokkum et al. 2004; Caputi et al. 2004; Caputi et al. 2005a,b). Certainly, we find other $\rm 24 \, \mu m$ massive galaxies for which the derived lifetimes $M/SFR$ are larger (up to $\sim$ 1 Gyr), indicating that these galaxies would be present since higher redshifts and that  star-formation histories  longer than a single burst lifetime are necessary to build the stellar mass present in them. Moreover, other massive galaxies with SFR below the limits of the $24 \, \mu {\rm m}$ survey analyzed here also exist at redshift $z \sim 2-3$ (Caputi et al. 2005c, in preparation).

  Interestingly, we note that the best-fit optical to near-IR SEDs for our $\rm 24 \, \mu m$ galaxies at redshifts $z \gsim 2$  correspond to different galaxy ages, many of which exceed the maximum $\sim$ 1 Gyr lifetime derived with the ratio $M/SFR$. This fact could be indicating  that the instantaneous star-formation rates derived from the $\rm 24 \, \mu m$ fluxes of these galaxies were not constant or larger in the past.  In order for these galaxies to have ages of  $\sim$ 1-3 Gyr (as those derived from the best-fit SEDs), their star-formation rates must have been {\em smaller} at some moment in the past. This is strongly suggesting that some  massive galaxies could be already in place at higher redshifts  and experience a non-continuous star-formation history, with temporary episodes of high star-formation activity (cf. also Papovich et al.~2005b). Those suffering such episodes at $z \sim 2-3$ would be detected as hyper luminous infrared galaxies at these redshifts  in mid-IR surveys.    If this is the case, the major episodes of star-formation at  $z \sim 2-3$ would not account for the complete assembly history of these massive galaxies.   A similar argument  has been given by Hammer et al.~(2005), who propose that the stellar mass growth of intermediate-mass galaxies at $z<1$ could proceed by successive star-formation episodes characterized by a LIRG phase.

  As we mentioned in Section~\ref{sec_redsh}, a substantial fraction of the $\rm 24 \, \mu m$ sources at high redshifts have the characteristic colours of ERGs, $(I_{775} - K_s) >4.0$ (Vega). Based on the best-fit optical to near-IR SEDs,  the large majority of the ERGs have been formed in short timescales (instantaneously or with characteristic times $\rm \tau \leq 1 \, Gyr$). For many of them, the best-fit age is larger than the SED-derived time-formation scale, i.e. age $\rm > 1 \, Gyr$. However, even when these characteristics would correspond to systems undergoing passive evolution,  the SED of most of these ERGs can only be properly fitted with considerable amounts of dust extinction. We find that the median of the colour excess in the $\rm 24 \, \mu m$ ERGs at $z>2$ is $E(B-V) \approx 0.30$, a value too high to consider that these galaxies could be passive systems even from the optical to near-IR point of view.

   It should be emphasized that the SFRs computed here assume that the mid-IR sources are purely star-forming. Another possibility is that at least part of the mid-IR emission of these sources is due to the presence of an AGN, in which case the real SFR would be lower than our estimated values. We note, however, that even when e.g. only one half of the mid-IR luminosities of the $z \sim 2-3$ galaxies analyzed here were due to star formation, the derived SFR would still be sufficiently high as to construct stellar masses of the order $\gsim 10^{10} M_{\odot}$ in a burst timescale.  The properties of these high-redshift massive mid-IR galaxies  resemble those of submillimetre galaxies, whose redshift distribution has a median of $z=2.2-2.4$ \cite[]{chap03,chap05}. Thus, it is expected  that the  $\rm 24 \, \mu m$ galaxy population at the depth of the GTO images contains at least a fraction of the submillimetre galaxy population (cf. Egami et al. 2004; Ivison et al. 2004).

    At lower redshifts $z \lsim 1$, the completeness limits of the samples allow us to determine that  star-formation spans a wide range of timescales, in agreement with results obtained from the analysis of ISO sources (Franceschini et al. 2003). Moreover, we find that  these  timescales for star-formation activity depend on galaxy mass, with more massive star-forming galaxies  probably experiencing more prolonged star-formation histories. In particular, for the high-mass end   $M \geq 1.5 \times 10^{11} M_\odot$ at $z \lsim 1$, we see that the derived timescales are larger than the age of the Universe at the corresponding redshifts. This fact indicates that the SFR of these galaxies was not constant over time, but larger in the past. This plausible differential star-formation history for star-forming galaxies of different assembled stellar masses  at  $z \lsim 1.0$ is in agreement with recent results reported in the literature (Heavens et al. 2004).

     Although the derived timescales $M/SFR$ for  massive ($M \geq 1.5 \times 10^{11} M_\odot$) galaxies at $z \lsim 1$  indicate prolonged star-formation histories, different possibilities exist for the way in which star formation proceeded through cosmic time. Star formation could have proceeded quiescently, i.e. with relatively small SFR sustained through long ($>0.1$ Gyr) periods of time. Alternatively, as we discussed before, star formation might have been non-continuous, with the production of multiple burst-like episodes. Both ways -or also some combination of the two- could have lead to the construction of the stellar mass contained in these massive galaxies. However, the number densities of both all and $S_\nu(24 \, \mu{\rm m})>83 \, \mu {\rm Jy}$ massive galaxies at $z \sim 3$ is  only $\sim$ 20\% of the corresponding number densities at $z<1$ (Caputi et al. 2005b,c). Thus, the major episodes of star-formation activity produced at very high redshifts could  be responsible for the construction of stellar mass  in at most a minor fraction of the massive $24 \, \mu \rm m$  galaxies observed at $z<1$.
 Also, it is likely that most of the massive galaxies undergoing high star-formation activity at $z \sim 2-3$ do not have significant amounts of  remaining gas to condense into stars by redshift $z \sim 1$. Thus, massive $24 \, \mu \rm m$  galaxies at  $z \sim 2-3$ are probably not the progenitors of the  massive $24 \, \mu \rm m$  galaxies at  $z \sim 1$. The star-formation history of massive IR galaxies
 at low redshifts should have mainly proceeded by some combination of quiescent activity and relatively modest burst-like episodes.  
     
     It is interesting to analyze the evolution with redshift of the efficiency of the burst-like star-formation mode to construct a significant fraction of a galaxy stellar mass. As we said above, at redshifts $z \sim 2-3$,  the typical starburst lifetime of 0.01 Gyr to 0.1 Gyr is sufficient to form a substantial fraction of the stellar mass of some massive galaxies ( $M \geq 1.5 \times 10^{11} M_\odot$). At redshifts $z \sim 1-2$, only low to intermediate-mass galaxies ($M \leq 7.0 \times 10^{10} M_\odot$) could form in a burst lifetime.  Within the limits of our sample and the volume surveyed, no galaxy appears to be able to have built up in a burst timescale at $z \lsim 1$. This does not mean that the burst-like mode cannot proceed in different mass galaxies at these low redshifts, but each individual burst would only be sufficient to construct an additional minor amount of the galaxy assembled stellar mass (i.e. the specific star formation rate $SFR/M$ would be low).  Thus, our results suggest that  the potential importance of the burst-like mode of formation shifts from high to low-mass galaxies with decreasing redshift (note, however, that we cannot completely probe star-formation in galaxies with stellar mass $M \lsim 10^{11} M_\odot$ at $z \sim 2-3$). In fact, each individual burst-like episode can play only a secondary role in galaxy building at $z \lsim 1$ (see also Papovich et al. 2005b).


\section{Discussion and summary} 
\label{sec_disc} 

  In this work, we presented $K_s$-band identifications of the MIPS $24 \mu {\rm m}$ galaxies in the GOODS/CDFS at the depth of the {\em Spitzer}/GTO surveys (with $\sim$ 94\% of identification completeness for sources with flux $S_\nu(24 \, \mu{\rm m})>83 \, \mu {\rm Jy}$). A $K_s<21.5$ (Vega) galaxy survey is sufficient for this purpose. The remaining non-identified galaxies are mostly faint mid and near-IR sources. A minor fraction (7\%) of the 747 identified objects correspond to  X-ray-selected active galaxies. Active galaxies are among the brightest $\rm 24 \mu {\rm m}$ sources at high redshifts (cf. Houck et al. 2005).

 MIPS $\rm 24 \mu m$ galaxies in the GOODS/CDFS span the redshift range $z=0-4$ and  28\% of the population lies at redshifts $z \gsim 1.5$ at the depth of the GTO data.  This confirms the predictions of a substantial population of mid-IR-selected  galaxies  at high redshifts. We determined the existence of a significant bump in the redshift distribution of $\rm 24 \mu m$ galaxies at $z \sim 1.9$, in agreement with predictions made by Lagache et al.~(2004) using the Lagache et al.~(2003) model. This secondary peak in the redshift distribution of  $\rm 24 \mu m$ galaxies indicates the existence of  PAH spectral features in $z \sim 1.9$ star-forming galaxies. We also observed the predicted depression in the redshift distribution at $z \sim 1.5$, although with marginal significance, which could be driven by silicate absorption. However, the overall observed redshift distribution differs significantly from the distribution predicted by the Lagache et al.(2003) and other models, indicating that revisions are required to understand the high-redshift infrared-luminous galaxy population.

  The mid-IR galaxy population is mainly composed of normal star-forming galaxies at redshifts $0.4 \leq z < 0.8$, while LIRGs start to be dominant  at $0.8 \leq z < 1.2$.  We derived estimated assembled stellar masses of $10^9 \, M_\odot$ to $10^{12} \, M_\odot$ for the $24 \, \mu {\rm m}$ galaxies at $z \lsim 1.2$. Massive star-forming galaxies appear to dominate the mid-IR output progressively at higher redshifts. A considerable fraction of the stellar mass density of the Universe is contained  in bright $24 \,\mu {\rm m}$ sources: galaxies with flux $S_\nu(24 \, \mu{\rm m})>83 \, \mu {\rm Jy}$ contain $\sim$ 35\% of the stellar mass density at redshifts $z=0.5-1.5$ and $\sim$ 40\% of the stellar mass density at $z \sim 1.5-3.0$. This corresponds to $\sim$ 30\%, 15\%, 10\% and a minimum of 5\% of the local value at redshifts $z \approx 0.75, 1.25, 1.75$ and $2.5$, respectively (Caputi et al. 2005c). Thus, we conclude that bright  $24 \,\mu {\rm m}$  galaxies have a significant role in the history of stellar mass assembly.

  We find that, at redshifts $z \lsim 1.2$, the IR-derived timescales for star-formation  activity increase with  the already-assembled galaxy mass. Mid-IR galaxies with intermediate to high assembled stellar masses  ($M \gsim 10^{10}  \, M_{\odot}$) appear to be  suffering rather prolonged star-formation histories, while lower-mass  ($M \lsim 10^{10}  \, M_{\odot}$) galaxies, on the contrary, are characterized by  star-formation activity on shorter  timescales.    At higher redshifts $z \gsim 2$, massive galaxies are dominant and all the derived timescales for the $S_\nu(24 \, \mu {\rm m}) >83 \, \mu {\rm Jy}$ sources are $\rm \lsim 1 \, Gyr$. Star-formation history at high redshifts is likely to have proceeded in a series of burst-like episodes and, in a substantial fraction of massive galaxies, a single  burst lifetime (0.01 Gyr - 0.1 Gyr) is sufficient to assemble a stellar mass of $\approx 10^{10}  \, M_{\odot} - 10^{11}  \, M_{\odot}$. We conclude, then, that the burst-like mode was probably an efficient  way to construct the bulk of  the stellar mass of massive galaxies  at high redshifts. 
 
  At $z \sim 1-2$, on the contrary,  the burst-like mode can be effective only for the formation of galaxies with mass  $M \lsim 7.0 \times 10^{10} M_\odot$ , and individual bursts can play only a secondary role in galaxy-building by redshifts  $z \lsim 1$. Recently, Bell et al.~(2005) found that the majority of the  $\rm 24 \, \mu m$ galaxies present at redshift $z \sim 0.7$ in the CDFS have regular spiral morphologies rather than irregular or peculiar morphologies. In conjunction, Bell et al. morphological determinations and our results on the lack of  galaxies built up in a burst at $z \lsim 1$ suggest that  morphology  could be linked to the mode of star-formation activity (see also Papovich et al. 2005b). Galaxies with no on-going  or mainly quiescent star-formation would show regular shapes, while galaxies suffering important burst-like episodes of star-formation activity would appear as irregular. To test this hypothesis, a detailed morphological study of  $\rm 24 \, \mu m$ galaxies as a function of stellar mass and redshift would be necessary.

   Results recently published in the literature suggest a `down-sizing' history for the formation of the stellar mass, where star-formation proceeds from high to low-mass galaxies (Heavens et al. 2004;  Juneau et al. 2005). Our conclusions support  a similar down-sizing mechanism for the efficiency of the burst-like mode.  This fact also suggests that massive star-forming galaxies at high redshifts are not the progenitors of the massive star-forming galaxies  found to be at $z \lsim 1$, which are characterized by much longer star-formation histories.    The properties of   $\rm 24 \, \mu m$ galaxies at $z \gsim 2$ suggest, on the contrary,  a likely connection with submillimetre galaxies. Current interpretations of submillimetre galaxies indicate that this population could contain the progenitors of the most massive E/S0 galaxies present in the local Universe (e.g. Stevens et al. 2003). The high SFR characterizing mid-IR-selected galaxies at $2 \lsim z \lsim 3$  suggests a similar conclusion for the high-redshift $\rm 24 \,\mu m$ galaxy population. However, some recent works suggest that a non-negligible fraction of present-day massive galaxies would be in place at even higher $z \gsim 3$ redshifts, indicating that the determination  of the first epoch of formation of  massive spheroids is still to be discovered.

%
\acknowledgments 
This paper is based on observations made with the {\it Spitzer}
Observatory, which is operated by the Jet Propulsion Laboratory, 
California Institute of Technology under NASA contract 1407. Also based on observations made with the Advanced Camera for Surveys  on board the Hubble Space Telescope operated by NASA/ESA and with the Infrared
Spectrometer and Array Camera on the `Antu' Very Large Telescope operated by the
European Southern Observatory in Cerro Paranal, Chile, and which form part of the publicly available GOODS datasets. We thank the GOODS teams for providing
reduced data products.

 We thank the anonymous referee for his/her useful comments and suggestions which improved the discussion of results in this paper.  Support for this work was provided by NASA 
through Contract Number 960785 issued by JPL/Caltech. KIC acknowledges CNES and CNRS funding. RJM acknowledges the support of the Royal Society.

%

\clearpage

%

%

%

\clearpage

%

\clearpage

%

%

%

%

%

%

%
\begin{deluxetable}{rc} 
\tablewidth{7cm} 
\tablecaption{Completeness limits for the $K_s$-band identifications of the {\it Spitzer} MIPS $\rm 24\, \mu m $ sources in the GOODS/CDFS.
\label{tab1}} 
\tablehead{ 
\colhead{24~$\mu$m flux} & 
\colhead{$K_s$-band identification} \\ 
\colhead{$S_\nu$ [$\mu$Jy]} &
 \colhead{completeness}
} 
\startdata 
$>250$ & 100\% \\ 
$>200$ & 99\% \\ 
$>150$ & 96\% \\
$>100$ & 95\% \\
$> \, \,\, 83$  & 94\% \\
\enddata 
\end{deluxetable}

\end{document}